\documentclass[aoas,MSNbibl,nameyear,rotating,dvips]{arximspdf}
\usepackage{dcolumn}
\usepackage{graphicx}

\doi{10.1214/12-AOAS592} 
\volume{7}
\issue{1}
\pubyear{2013}
\firstpage{418}
\lastpage{442}

\makeatletter

\newcolumntype{d}[1]{D{.}{.}{#1}}

\newproclaim{Remark}{Remark}

\newtheorem{corollary}{Corollary}

\newtheorem{theorem}{Theorem}

\newcommand{\rrVert}{\Vert}
\newcommand{\llVert}{\Vert}

\newcommand{\nulll}{{\mathbf{0}}}
\renewcommand{\b}{{\mathbf{b}}}
\renewcommand{\d}{{\mathbf{d}}}
\newcommand{\f}{{\mathbf{f}}}
\renewcommand{\u}{{\mathbf{u}}}
\renewcommand{\H}{{\mathbf{H}}}
\newcommand{\I}{{\mathbf{I}}}
\newcommand{\X}{{\mathbf{X}}}
\newcommand{\Y}{{\mathbf{Y}}}

\newcommand{\balpha}{\bolds{\alpha}}
\newcommand{\bbeta}{\bolds{\beta}}
\newcommand{\bphi}{\bolds{\phi}}
\newcommand{\bgamma}{\bolds\gamma}

\newcommand{\diag}{\operatorname{diag}}
\newcommand{\reals}{\mathbb{R}}

\newcommand{\bsk}{_{\bar{S}k}}

\newcommand{\var}{\operatorname{Var}}

\newcommand{\E}{\mathrm{E}}
\newcommand{\argmin}{\mathop{\arg\min}}

\makeatother

\begin{document}
\begin{frontmatter}

\title{Variable selection for sparse Dirichlet-multinomial regression
with an application to microbiome
data~analysis\thanksref{T1}}
\runtitle{Sparse Dirichlet-multinomial regression}

\thankstext{T1}{Supported in part by NIH Grants CA127334, GM097505 and
UH2DK083981.}

\begin{aug}
\author[A]{\fnms{Jun} \snm{Chen}\ead[label=e1]{chenjun@mail.med.upenn.edu}}
\and
\author[A]{\fnms{Hongzhe} \snm{Li}\corref{}\ead[label=e2]{hongzhe@upenn.edu}}
\runauthor{J. Chen and H. Li}
\affiliation{University of Pennsylvania}
\address[A]{Department of Biostatistics\\
\quad and Epidemiology\\
University of Pennsylvania\\
Philadelphia, Pennsylvania 19104-6021\\
USA\\
\printead{e1}\\
\hphantom{E-mail: }\printead*{e2}} 
\end{aug}

\received{\smonth{11} \syear{2011}}
\revised{\smonth{8} \syear{2012}}

%
\begin{abstract}
With the development of next generation sequencing technology,
researchers have now been able to study the microbiome composition
using direct sequencing, whose output are bacterial taxa counts for
each microbiome sample. One goal of microbiome study is to associate
the microbiome composition with environmental covariates. We propose
to model the taxa counts using a Dirichlet-multinomial (DM) regression
model in order to account for overdispersion of observed counts. The
DM regression model can be used for testing the association between
taxa composition and covariates using the likelihood ratio test.
However, when the number of covariates is large, multiple testing can
lead to loss of power. To address the high dimensionality of the
problem, we develop a penalized likelihood approach to estimate the
regression parameters and to select the variables by imposing a sparse
group $\ell_1$ penalty to encourage both group-level and within-group
sparsity. Such a variable selection procedure can lead to selection
of the relevant covariates and their associated bacterial taxa. An
efficient block-coordinate descent algorithm is developed to solve the
optimization problem. We present extensive simulations to demonstrate
that the sparse DM regression can result in better identification of
the microbiome-associated covariates than models that ignore
overdispersion or only consider the proportions. We demonstrate the
power of our method in an analysis of a data set evaluating the
effects of nutrient intake on human gut microbiome composition. Our
results have clearly shown that the nutrient intake is strongly
associated with the human gut microbiome.
\end{abstract}

%

\end{frontmatter}

\section{Introduction}
The human body is inhabited by complex microbial communities, called
microbiomes. It is estimated that the number of microbial cells
associated with the human body is 10 times the total number of human
cells. The collective genomes of these microbes constitute an extended
human genome that provides us with genetic and metabolic capabilities
that we do not inherently possess [\citet{Backhed2005}]. With the
development of next generation sequencing technology such as the 454
pyrosequencing and Illumina Solexa sequencing, microbiome composition
can now be determined by direct DNA sequencing without laborious
cultivation. Typically, instead of sequencing all bacterial genomic DNA
as in a shotgun metagenomic approach, only the 16S rRNA gene, which is
ubiquitous in the bacteria kingdom and has variable regions, is
sequenced. Since each bacterial cell is assumed to have the same number
of copies of this gene, the basic idea is to isolate from all the
bacteria the DNA strands corresponding to some variable region of the
gene, to count different versions of the sequences, and then to
identify to which bacteria the versions correspond. The types and
abundances of different bacteria in a sample can therefore be
determined. After preprocessing of the raw sequences, the 16S sequences
are either mapped to an existing phylogenetic tree in a taxonomic
dependent way [e.g., \citet{Matsen2010}] or clustered into operational
taxonomic units (OTUs) at a certain similarity level in a taxonomic
independent way [e.g., \citet{Caporaso2010a,Schloss2009a}]. At 97\%
similarity level, these OTUs are used to approximate the taxonomic rank
\textit{species}. The OTU based approach is most commonly used in 16S
based microbiome studies. Each OTU is characterized by a
representative DNA sequence and can be assigned a taxonomic lineage by
comparing to a known bacterial 16S rRNA database. Most OTUs are in
extremely low abundances, with a large proportion being simply
singletons (possibly due to sequencing error). We can further aggregate
OTUs from the same genus and perform analysis on the abundances at the
genus level, which is more robust to sequencing error and can reduce
the number of variables significantly. Either way we finally obtain the
taxa counts for each sample.

Recent studies have linked the microbiome with human diseases
including \mbox{obesity} and inflammatory bowel disease [\citet{Virgin2011}].
It is therefore important to understand how genetic or environmental
factors shape the human microbiome in order to gain insight into
etiology of many microbiome-related diseases and to develop therapeutic
measures to modulate the microbiome composition. \citet{Benson2010}
demonstrated that genetic variants are associated with the mouse gut microbiome.
\citet{Wu2011a} showed that dietary nutrients are associated with the
human gut microbiome.
Both studies have considered a large number of genetic loci or
nutrients and aimed to identify the genetic variants or nutrients that
are associated with the gut microbiome. When there are numerous
possible covariates affecting the microbiome composition, variable
selection becomes necessary. Variable selection cannot only increase
biological interpretability but also provide researchers with a short
list of top candidates for biological validation. The methods we\vadjust{\goodbreak}
develop in this paper are particularly motivated by an ongoing study at
the University of Pennsylvania to link the nutrient intake to the human
gut microbiome. In this study, gut microbiome data were collected on 98
normal volunteers. In addition, food frequency questionnaire (FFQ) were
filled out by these individuals. The questionnaires were scored and the
quantitative measurements of 214~micronutrients were obtained. Details
of the study and the data set can be found in Section~\ref{secreal}
and in \citet{Wu2011a}. Our goal is to identify the nutrients that are
associated with the gut microbiome and also their associated bacterial taxa.

Most of the microbiome studies used distance-based methods
to link the microbiome and environmental covariates, where a distance
metric was defined between two microbiome samples and statistical
analysis was then performed using the distances. However, the choice of
distance metric is sometimes subjective and different distances vary in
their power of identifying relevant environmental factors. Another
limitation of distance-based methods is its inefficiency for detecting
subtle changes since distances summarize the overall relationship. In
addition, such distance-based approaches do not provide information on
how covariates affect the microbiome compositions and which taxa are
affected. Therefore, it is desirable to model the counts directly
instead of summarizing the data as distances.
One way of testing for covariate effects is by performing a
multivariate multiple regression (called redundancy analysis in
ecology) after appropriate transformation of the count data such as
converting into proportions [\citet{PierreLegendre2002}]. A pseudo-$F$
statistic is then calculated and the significance is then evaluated by
permutation test. Alternatively, one can define a distance between the
samples and then use a PERMANOVA procedure to test for covariate
effects [\citet{McArdle2001}]. It is easy to show that when the distance
is Euclidean, these two procedures are equivalent.

In this paper, we consider the sparse Dirichlet-multinomial (DM)
regression [\citet{MOSIMANN1962}] to link high-dimensional covariates to
bacterial taxa counts from microbiome data. The DM regression model is
chosen to model the overdispersed taxa counts. The observed taxa count
variance is much larger than that predicted by a multinomial model that
assumes fixed underlying taxa proportions, an assumption that is hardly
met for real microbiome data. Uncontrollable sources of variation such
as individual-to-individual variability, day-to-day variability,
sampling location variability or even technical variability such as
sample preparation lead to enormous variability
in the underlying proportions. In contrast, the DM model assumes that
the underlying taxa proportions come from a Dirichlet distribution. We
use a log-linear link function to associate the mean taxa proportions
with covariates. In this DM modeling framework, the effects of the
covariates on taxa proportions can be tested using the likelihood ratio test.

When the number of the covariates is large, we propose a sparse group
$\ell_1$ penalized likelihood approach for variable selection and
parameter estimation. The\vadjust{\goodbreak} sparse group $\ell_1$ penalty function
[\citet{Friedman2010}] consists of a group $\ell_1$ penalty and an
overall $\ell_1$ penalty, which induce both group-level sparsity and
within-group sparsity. This is particularly relevant in our setting.
For the nutrient-microbiome association example, we have $p$ nutrients
and $q$ taxa, so the fully parameterized model has $(p+1)\times q$
coefficients including the intercepts, since each nutrient-taxon
association is characterized by one coefficient. The $q$ coefficients
for each nutrient constitute a group. If we assume many nutrients have
no or ignorable effects on the microbiome composition, the groups of
coefficients associated with these irrelevant nutrients should be zero
altogether, which is a group-level sparsity that is achieved by
imposing a group $\ell_1$ penalty. However, the group $\ell_1$ penalty
does not perform within-group selection, wherein if one group is
selected, all the coefficients in that group are nonzeros. In the case
of nutrient-microbiome association, we are also interested in knowing
which taxa are associated with a selected nutrient. By imposing an
overall $\ell_1$ penalty, within-group selection becomes possible.
Therefore, we impose a sparse group $\ell_1$ penalty not only to select
these important nutrients but also to recover relevant nutrient-taxon
associations.

Section~\ref{secDM} reviews the Dirichlet-multinomial model for count
data. Section~\ref{secLR} introduces the Dirichlet-multinomial
regression framework for incorporating covariate effects and proposes a
likelihood ratio statistic for testing the covariate effect. Section
\ref{secgrp} proposes a sparse group $\ell_1$ penalized likelihood
procedure for variable selection for the DM models followed by a
detailed description of a block-coordinate descent algorithm in Section
\ref{secalg}. Section~\ref{secsim} shows simulation results and
Section~\ref{secreal} demonstrates the proposed method on a real human
gut microbiome data set to associate the nutrient intake with the human
gut microbiome composition.

\section{Dirichlet-multinomial model for microbiome composition data}
\label{secDM}
Suppose we have $q$ bacterial taxa and their counts $Y=(Y_1, Y_2,\ldots,
Y_q)$ are random variables. Denote ${\mathbf{y}}=(y_1,y_2,\ldots,y_q)$ as the
observed counts. The simplest model for count data is the multinomial
model and its probability function is given as
\[
f_M(y_1, y_2, \ldots, y_q;
\bphi) = \pmatrix{y_+ \cr{\mathbf{y}}}\prod_{j=1}^{q}
\phi_j^{y_j},
\]
where $y_+=\sum_{j=1}^q y_j$ and $\bphi=(\phi_1, \phi_2, \ldots,
\phi_q)$ are underlying species proportions with $\sum_{j=1}^q\phi_j=1$.
Here the total taxa count $y_+$ is determined by the sequencing depth
and is treated as an ancillary statistic since its distribution does
not depend on the parameters in the model. The mean and variance of the
multinomial component~$Y_j$ $(j=1 ,\ldots, q)$ are
%
\begin{equation}
\label{eqM2} \E(Y_j) = y_+\phi_j,\qquad
\var(Y_j) = y_+ \phi_j(1-\phi_j).
\end{equation}

For microbiome composition data, the actual variation is usually larger
than what would be predicted by the multinomial model,\vadjust{\goodbreak} which assumes
fixed underlying proportions. This increased variation is due to the
heterogeneity of the microbiome samples and the underlying proportions
vary among samples. To account for the extra variation or
overdispersion, we assume the underlying proportions $(\phi_1, \phi_2,
\ldots, \phi_q)$ are themselves positive random variables $(\Phi_1,
\Phi_2, \ldots, \Phi_q)$ subject to the constraint $\sum_{j=1}^q \Phi_j
=1$. One commonly used distribution is the Dirichlet distribution
[\citet{MOSIMANN1962}] with the probability function given by
\[
f_D(\phi_1, \phi_2, \ldots,
\phi_q; \bgamma) = \frac{\Gamma
(\gamma_+)}{\prod_{j=1}^q \Gamma(\gamma_j)} \prod
_{j=1}^q \phi_j^{\gamma
_j -
1},
\]
where $\bgamma=(\gamma_1,\gamma_2,\ldots,\gamma_q)$ are positive
parameters, $\gamma_+=\sum_{j=1}^q \gamma_j$ and $\Gamma(\cdot)$
is the
Gamma function. The mean and variance of the Dirichlet component $\Phi_j$
$(j=1 ,\ldots, q)$ are
\[
\E(\Phi_j) = \frac{\gamma_j}{\gamma_+},\qquad \var(\Phi_j)=
\frac
{\gamma_j(\gamma_+ - \gamma_j)}{(1+\gamma_+)\gamma_+^2}.
\]
The mean is proportional to $\gamma_j$ and the variance is controlled
by $\gamma_+$, which can be regarded as a ``precision parameter.'' As
$\gamma_+$ becomes larger, the proportions are more concentrated around
the means.

The Dirichlet-multinomial (DM) distribution [\citet{MOSIMANN1962}]
results from a compound multinomial distribution
with weights from the Dirichlet distribution (parametrization I):
%
\begin{eqnarray}
\label{eqDM1} f_{\mathrm{DM}}(y_1, y_2, \ldots,
y_q;\bgamma) & = & \int f_M(y_1,
y_2, \ldots, y_q; \bphi)f_D(\bphi;\bgamma)\,d
\bphi
\nonumber\\[-8pt]\\[-8pt]
& = & \pmatrix{y_+ \cr{\mathbf{y}}}\frac{\Gamma(y_+ + 1) \Gamma(\gamma_+)}{\Gamma(y_+
+ \gamma_+)} \prod
_{j=1}^{q} \frac{\Gamma(y_j + \gamma_j)}{\Gamma
(\gamma_j)\Gamma(y_j+1)}.\nonumber
\end{eqnarray}
The mean and variance of the DM distribution for each component $Y_j$
$(j=1,\ldots,q)$ is given by
%
\begin{equation}
\label{eqDM2} \E(Y_j) = y_+\E(\Phi_j),\qquad
\var(Y_j) = y_+\E(\Phi_j)\bigl\{1 - \E (
\Phi_j)\bigr\} \biggl( \frac{y_+ + \gamma_+}{1 + \gamma_+} \biggr).
\end{equation}
Comparing (\ref{eqDM2}) with (\ref{eqM2}), we see that the variation
of the DM component is increased by a factor of $(y_+ + \gamma_+)/(1 +
\gamma_+)$, where $\gamma_+$ controls the degree of overdispersion with
a larger value indicating less overdispersion. Using an alternative
parameterization, the probability function can be written as
(parameterization~II)
%
\begin{equation}
\label{eqDM3} f_{\mathrm{DM}}^*(y_1, y_2,\ldots,
y_q; \bphi, \theta) = \pmatrix{y_+ \cr{\mathbf {y}}}\frac
{\prod_{j=1}^q\prod_{k=1}^{y_j}\{\phi_j(1-\theta) + (k-1)\theta\}
}{\prod_{k=1}^{y_+}\{1 - \theta+ (k-1)\theta\}},
\end{equation}
where $\phi_j= \gamma_j/\gamma_+$ is the mean and $\theta
=1/(1+\gamma_+)$ is the dispersion parameter. When $\theta= 0$, it is easy to
verify (\ref{eqDM3}) is reduced to the multinomial distribution.

\section{Dirichlet-multinomial regression for incorporating the
covariate effects} \label{secLR}
When there is no covariate effect, the DM model can be used to produce
more accurate estimates of taxa proportions of a given microbiome
sample than the simple multinomial model, due to its ability to model
the overdispersion. Beyond proportion estimation, microbial ecologists
are more interested in associating the microbiome composition with some
environmental covariates.
Suppose we have $n$ microbiome samples and $q$ species. Let $\Y
=(y_{ij})_{n \times q}$ be the observed count matrix for the $n$
samples. Let $\X=(x_{ij})_{n \times p}$ be the design matrix of $p$
covariates for $n$ samples. We assume the parameters $\gamma_j$ $(j=1,\ldots,q)$ in the DM model (parametrization I) depend on the
covariate via the following log-linear model,
%
\begin{equation}
\label{eqlink1} \gamma_j\bigl({\mathbf{x}}^i\bigr) =
\exp\Biggl(\alpha_j + \sum_{k=1}^p
\beta_{jk}x_{ik}\Biggr),
\end{equation}
where ${\mathbf{x}}^i$ is the $i$th row vector of $\X$ and $\beta_{jk}$ is the
coefficient for the $j$th taxon with respect to $k$th covariate, whose
sign and magnitude measure the effect of the $k$th covariate on the
$j$th taxon.
From (\ref{eqDM2}), we see that $\E(Y_{ij}) \propto\exp(\alpha_j)\prod_{k=1}^p\exp(\beta_{jk}x_{ik})$, where $\exp(\alpha_j)$ can be
interpreted as the baseline abundance level for species $j$ and the
coefficient $\beta_{jk}$ indicates the magnitude of the $k$th covariate
effect on species $j$. Though the log-linear link is assumed mainly for
ease of computation, it is biologically consistent, in that
microorganisms usually
exhibit exponential growth in a favorable environment.\looseness=-1

For notational simplicity, we denote $\beta_{j0}$ as $\alpha_j$ and
augment $\X$ with an $n$-vector of $1$'s as its first column. We number
the columns from $0$ to $p$. The link function becomes
%
\begin{equation}
\label{eqlink2} \gamma_j\bigl({\mathbf{x}}^i\bigr) =
\exp\Biggl(\sum_{k=0}^p
\beta_{jk}x_{ik}\Biggr).
\end{equation}
Let $\bbeta$ be the $q\times(p+1)$ regression coefficient matrix,
$\bbeta^{j}=(\beta_{j0},\ldots,\beta_{jp})^T$ be the vector of
coefficients for the $j$th taxon ($j=1,\ldots,q$) and $\bbeta_{k}=(\beta_{1k},\ldots,\break\beta_{qk})^T$ be the vector of coefficients for the $k$th
covariate ($k=0,\ldots,p$). We also use $\bbeta$ to denote the
$q(p+1)$ vector that contains all the coefficients. Substituting (\ref
{eqlink1}) into DM probability function (\ref{eqDM1}) and ignoring the
part that does not involve the parameters, the log-likelihood function
given the covariates is given by
%
\begin{eqnarray}
\label{eqloglik2} l(\bbeta; \Y,\X) &=& \sum_{i=1}^n
\Biggl[ \tilde{\Gamma} \Biggl(\sum_{j=1}^q
\gamma_j\bigl({\mathbf{x}}^i;\bbeta^j\bigr)
\Biggr) - \tilde{\Gamma } \Biggl(\sum_{j=1}^q
y_{ij} + \sum_{j=1}^q
\gamma_j\bigl({\mathbf{x}}^i;\bbeta^j\bigr)
\Biggr)
\nonumber\\[-9pt]\\[-9pt]
&&\hspace*{45pt}{} + \sum_{j=1}^q \bigl\{\tilde{
\Gamma} \bigl(y_{ij}+\gamma_j\bigl({\mathbf{x}}^i;
\bbeta^j\bigr) \bigr) - \tilde{\Gamma } \bigl(\gamma_j
\bigl({\mathbf{x}}^i;\bbeta^j\bigr) \bigr) \bigr\}
\Biggr],\nonumber
\end{eqnarray}
where $\tilde{\Gamma}(\cdot)$ is the log-gamma function.\vadjust{\goodbreak}

Based on the likelihood function (\ref{eqloglik2}), one can test the
effect of a given covariate or the joint effects of all covariates on
the microbiome
composition using the standard likelihood ratio test (LRT). To solve
the maximization problem, we implemented the Newton--Raphson algorithm,
since the gradient and Hessian matrix of the log-likelihood can be
calculated analytically. Alternatively, we can use the general-purpose
optimization algorithm such as $nlm$ in R, which computes the gradient
and Hessian numerically. By selecting an appropriate starting point
(e.g., $\balpha=\bbeta=\nulll$), for moderate-size
problems in the dimensions $p$ and $q$, the algorithm converges to a
stationary point sufficiently fast.

With a large number of covariates in the DM regression model, direct
maximization of the likelihood function becomes infeasible or unstable.
When each covariate is tested separately using the LRT, adjustment for
multiple testing is required. In addition, when the number of taxa $q$
is large, the null distribution of the LRT has large degrees of freedom
and therefore reduced power. It is also desirable to select the
relevant covariates that are associated with the microbiome
composition. Although one can test the null hypothesis
$H_0\dvtx\beta_{jk}=0$ for each $(j,k)$ pair by the LRT, adjustment of multiple
comparisons can lead to a loss of power. In the next section we present
a sparse group $\ell_1$ penalized estimation for variable selection and
parameter estimation for sparse DM regression models.

\section{Variable selection for sparse Dirichlet-multinomial
regression} \label{secgrp}

To perform variable selection, we estimate the regression coefficient
vector $\bbeta$ in model (\ref{eqlink2}) by minimizing the following
sparse group $\ell_1$ penalized negative log-likelihood function,
%
\begin{equation}
\label{eqploglik} \mathit{pl}(\bbeta; \Y,\X, \lambda_1,
\lambda_2) = - l(\bbeta; \Y,\X) + \lambda_1 \sum
_{k=1}^p \llVert \bbeta_{k}
\rrVert_2 + \lambda_2 \sum_{k=1}^p
\llVert \bbeta_{k} \rrVert_1,
\end{equation}
where $l(\bbeta; \Y,\X)$ is the log-likelihood function defined as in
(\ref{eqloglik2}), $\lambda_1$ and $\lambda_2$ are the tuning
parameters and $\llVert \bbeta_k \rrVert_1=\sum_{j=1}^q|\beta_{ik}|$ is the
$\ell_1$ norm and $\llVert \bbeta_k \rrVert_2=\sqrt{\sum_{j=1}^q \beta_{ik}^2}$
is the group $\ell_1$ norm of the coefficient vector
$\bbeta_k$, respectively. We do not penalize the intercept vector
$\bbeta_0$. The first part of the sparse group $\ell_1$ penalty is the
group $\ell_1$ penalty that induces group-level sparsity, which
facilitates selection of the covariates that are associated with taxa
proportions. The second $\ell_1$ penalty on all the coefficients
facilitates the within-group selection, which is important for
interpretability of the resulting model. A similar penalty involving
both group $\ell_1$ and $\ell_1$
terms is discussed in \citet{Peng2009} and \citet{Friedman2010} for
regularized multivariate linear regression. When
$\lambda_2=0$, criterion (\ref{eqploglik}) reduces to the group lasso.

\subsection{A block-coordinate gradient descent algorithm for sparse
group $\ell_1$ penalized DM regression} \label{secalg}
The sparse group $\ell_1$ estimates of $\bbeta$ can be obtained by
minimizing the penalized negative log-likelihood function (\ref{eqploglik}):
\[
\hat{\bbeta}_{\lambda_1,\lambda_2} = \argmin_{\bbeta} \Biggl\{- l(\bbeta; \Y,\X)
+ \lambda_1 \sum_{k=1}^p
\llVert \bbeta_k \rrVert_2 + \lambda_2 \sum
_{k=1}^p \llVert \bbeta_k
\rrVert_1 \Biggr\}.
\]
Using the general block coordinate gradient descent algorithm of
\citet{Tseng2007}, we develop in the following an efficient
algorithm to solve this optimization problem. \citet{Meier2008}
present a block coordinate gradient descent algorithm for group lasso
for logistic regression that includes only the group $\ell_1$ penalty
(i.e., $\lambda_2=0$). In contrast, our optimization problem
(\ref{eqploglik}) has two nondifferentiable parts, both at the
individual $\beta_{jk}$ and at the group $\bbeta_k$ levels.

The key idea of the algorithm is to combine a quadratic approximation
of the log-likelihood function with an additional line search. First we
expand (\ref{eqloglik2}) at current estimate $\hat{\bbeta}^{(t)}$
to a
second-order Taylor series. The Hessian matrix is then replaced by a
suitable matrix $\H^{(t)}$. We define
%
\begin{equation}
\label{Taylor} l_Q^{(t)}(\d) = l\bigl(\hat{
\bbeta}{}^{(t)}\bigr) + \d^T\nabla l\bigl(\hat {
\bbeta}{}^{(t)}\bigr) + \tfrac{1}{2}\d^T\H^{(t)}
\d,
\end{equation}
where $\d\in\reals^{q(p+1)}$. Also denote $\nabla l(\hat{\bbeta
}{}^{(t)})_k$ and $\d_k$ the gradient and increment with respect to
$\hat
{\bbeta}{}^{(t)}_k$ for the $k$th group, and $\nabla l(\hat{\bbeta
}{}^{(t)})_{sk}$ and $\d_{sk}$ with respect to $\hat{\beta}{}^{(t)}_{sk}$.
We then minimize the following function $\mathit{pl}_{Q}^{(t)}(\d)$ with respect
to the $k$th penalized parameter group:
%
\begin{eqnarray}
\label{eqpqlik} \mathit{pl}_{Q}^{(t)}(\d) & = & - l_Q^{(t)}(
\d) + \lambda_1\sum_{k=1}^{p}
\bigl\llVert \hat{\bbeta}{}^{(t)}_k+\d_k \bigr
\rrVert_2 +\lambda_2\sum_{k=1}^{p}
\bigl\llVert \hat{\bbeta}{}^{(t)}_k+\d_k \bigr
\rrVert_1
\nonumber\\[-8pt]\\[-8pt]
&\approx& \mathit{pl}\bigl(\hat{\bbeta}{}^{(t)}+\d; \Y, \X, \lambda_1,
\lambda_2\bigr).
\nonumber
\end{eqnarray}
We restrict ourselves to vectors $\d$ with $\d_j = \nulll$ for $j
\neq k$
and the corresponding $q \times q$ submatrix $\H_{kk}^{(t)}$ for the
$k$th group is a diagonal matrix of the form $\H_{kk}^{(t)}=h_k^{(t)}\I_q$ for some scalar $h_k^{(t)} \in\reals$.

The solution\vspace*{1pt} to the general optimization problem of the form (\ref
{eqpqlik}) is given by Theorem~\ref{theorem1} and its corollary in the
\hyperref[app]{Appendix}. Let $S = \{s \vert \vert\nabla l(\hat{\bbeta
}{}^{(t)})_{sk} - h_k^{(t)}\hat{\beta}{}^{(t)}_{sk} \vert< \lambda_2 \}$
and $\bar{S}$ be the set $\{1,\ldots,q\}\setminus S$. Denote $\d_{Sk}$
the subvector of $\d_k$ with indices in $S$ and $\d\bsk$ in $\bar{S}$.
The minimizer of (\ref{eqpqlik}) can be decomposed into two parts:
The first part $\d_{Sk}^{(t)}$ can be obtained by
\[
\d_{Sk}^{(t)}= - \hat{\bbeta}{}^{(t)}_{Sk}.
\]
The second part $\d\bsk^{(t)}$ can be computed by minimizing
%
\begin{equation}
\label{eqpqlik2} f^{(t)}(\d_k) = - \bigl\{
\d_k^T \u_k^{(t)} +
\tfrac{1}{2}\d_k^T\H_{kk}^{(t)}
\d_k \bigr\} + \lambda_1 \bigl\llVert \hat{
\bbeta}{}^{(t)}_k+\d_k \bigr\rrVert_2
\end{equation}
with respect to $\d\bsk$ (set components other than $\d\bsk$ to be
$0$), where
\[
\u^{(t)}_k = \bigl[\nabla l\bigl(\hat{
\bbeta}{}^{(t)}\bigr)_k - \lambda_2
\operatorname{sgn} \bigl\{\nabla l\bigl(\hat{\bbeta}{}^{(t)}
\bigr)_k- h_k^{(t)}\hat{\bbeta
}{}^{(t)}_k \bigr\} \bigr]
\]
and sgn($\cdot$) is the sign function.

Minimization of (\ref{eqpqlik2}) with respective to $\d\bsk$ can be
performed in a similar fashion as in
\citet{Meier2008} for the group $\ell_1$ penalty. Specifically, if
$\llVert \u^{(t)}\bsk-h_k^{(t)}\bbeta^{(t)}\bsk\rrVert_2 <
\lambda_1$, the
minimizer of
equation (\ref{eqpqlik2}) for $\d\bsk$ is
\[
\d\bsk^{(t)}= -\hat{\bbeta}{}^{(t)}\bsk.
\]
Otherwise
\[
\d\bsk^{(t)}= -\frac{1}{h_k^{(t)}} \biggl\{\u^{(t)}\bsk-
\lambda_1 \frac{\u^{(t)}
\bsk-h_k^{(t)}\hat{\bbeta}{}^{(t)}\bsk}{
\llVert \u^{(t)}\bsk-h_k^{(t)}\hat{\bbeta}{}^{(t)}\bsk\rrVert_2} \biggr\}.
\]

For the unpenalized intercept, the solution can be directly computed:
\[
\d_0^{(t)} = -\frac{1}{h_0^{(t)}}\nabla l\bigl(\hat{
\bbeta}{}^{(t)}\bigr)_0.
\]

If $\d^{(t)}\ne\nulll$, an inexact line search using the Armijo
rule will
be performed. Let $\alpha^{(t)}$ be the largest value in $\{\alpha_0\delta^l\}_{l \ge0}$ such that
\[
\mathit{pl}\bigl(\hat{\bbeta}{}^{(t)}+\alpha^{(t)}\d^{(t)}
\bigr)-\mathit{pl}\bigl(\hat{\bbeta }{}^{(t)}\bigr) \le \alpha^{(t)}\sigma
\Delta^{(t)},
\]
where $ 0 < \delta<1, 0 < \sigma< 1, \alpha_0 > 0$, and $\Delta^{(t)}$
is the improvement in the objective function $\mathit{pl}(\bbeta)$
using a linear approximation, that is,
\begin{eqnarray*}
\Delta^{(t)}&=&-\d^{(t)T}\nabla l\bigl(\hat{\bbeta}{}^{(t)}
\bigr) + \lambda_1 \Biggl\{ \sum_{k=1}^p
\bigl\llVert \hat{\bbeta}{}^{(t)}_k + \d_k^{(t)}
\bigr\rrVert_2  -  \sum_{k=1}^p
\bigl\llVert \hat{\bbeta}{}^{(t)}_k \bigr\rrVert_2
\Biggr\}
\\
&&{}+\lambda_2 \Biggl\{\sum_{k=1}^p
\bigl\llVert \hat{\bbeta}{}^{(t)}_k + \d_k^{(t)}
\bigr\rrVert_1  -  \sum_{k=1}^p
\bigl\llVert \hat{\bbeta}{}^{(t)}_k \bigr\rrVert_1
\Biggr\}.
\end{eqnarray*}
Finally, we update the current estimate by
\[
\hat{\bbeta}{}^{(t+1)}=\hat{\bbeta}{}^{(t)} + \alpha^{(t)}
\d^{(t)}.
\]

For $\H_{kk}^{(t)}$, we use the same choice as in \citet{Meier2008},
that is,
\[
h_k^{(t)} = -\max \bigl[\diag\bigl\{-\nabla^2l
\bigl(\hat{\bbeta }{}^{(t)}\bigr)_{kk}\bigr\}, c^* \bigr],
\]
where $c^*>0$ is a lower bound to ensure convergence. In this paper, we
use the standard choices for the parameters, $\alpha_0=1,\delta
=0.5,\sigma=0.1$ and $c^*=0.001$ [\citet{Tseng2007}], in the block
coordinate descent algorithm to ensure
the convergence of the algorithm.

\begin{Remark*}
In each iteration of the algorithm detailed above, when estimating the
$k$th column of the $q\times p$ coefficient matrix $\bbeta$ with all
other columns fixed, the algorithm first identifies the coefficients
with zero estimates, denoted by set $S$ in the algorithm. For the
coefficients in set $S$, $d_{Sk}^{(t)}=-\hat{\bbeta}{}^{(t)}_{Sk}$ and,
therefore, when \mbox{$\alpha^{t}=1$},
$\hat{\bbeta}{}^{(t+1)}_{Sk}=\hat{\bbeta}{}^{(t)}_{Sk}+\alpha^{t}d_{Sk}^{(t)}=0$
and the coefficients\vspace*{1pt} in $S$ are shrunk to zero. Based on its
definition, the set $S$ depends on the turning parameter $\lambda_2$
and a larger value of $\lambda_2$ leads to fewer nonzero coefficients.
The algorithm then performs a group shrinkage of the
nonzero estimates of the coefficients in the complementary set
$\bar{S}$. These nonzero coefficients can further be shrunk to zero as
a group if the condition $\|\u^{(t)}\bsk-h_k^{(t)}\bbeta^{(t)}\|_2 <
\lambda_1$ is met, in which case
$d_{\bar{S}k}^{(t)}=-\hat{\bbeta}_{\bar {S}k}^{(t)}$ and, therefore,
$\hat{\bbeta}_{\bar{S}k}^{(t+1)}=\hat{\bbeta}_{{\bar
{S}k}}^{(t)}+d_{\bar {S}k}^{(t)}=0$. Clearly,\vspace*{2pt} this group
shrinkage depends on the tuning parameter $\lambda_1$. Thus, with
careful choice of the tuning parameters $\lambda_1$ and $\lambda_2$,
some column group coefficients are set to zero and the within-group
sparsity is achieved by the plain $\ell_1$ penalty.
\end{Remark*}

\subsection{Tuning parameter selection}

Two tuning parameters $\lambda_1$ and $\lambda_2$ in the penalized
likelihood estimation need to be tuned with data by $v$-fold
cross-validation or a BIC criterion. To facilitate computation, we
reparameterize $\lambda_1$ and $\lambda_2$ as
$\lambda_1=c\lambda\sqrt{q}$ and $\lambda_2=(1-c)\lambda$. The
multiplier $\sqrt{q}$ in the group penalty is used so that the group
$\ell_1$ penalty and overall $\ell_1$ penalty are on a similar scale.
Here we use $\lambda$ to control the overall sparsity level and use $c
\in[0,1]$ to control the proportion of group $\ell_1$ in the composite
sparse group penalty. When $c=0$, the penalty is reduced to the lasso;
when $c=1$, it is reduced to a group lasso. We consider the tuning
parameter $c$ from the set $\{0, 0.05, 0.1, 0.2, 0.4\}$. For each $c$,
to search for the best tuning parameter value, we run the algorithm
from $\lambda_{\mathrm{max}}$ so that it produces the sparsest model
with the intercepts $\bbeta_0$ only. The value $\lambda_{\mathrm{max}}$
can be roughly determined by using the starting value $\bbeta^{(0)}$
with components $\bbeta_j^{(0)}=\nulll$ $(j \neq0)$ and
$\bbeta_0^{(0)}$ the MLE of (\ref{eqloglik2}) without covariates, and
choosing the smallest value of $\lambda$ so that the iteration
converges in the first iteration, that is, $\bbeta^{(0)}$ is a
stationary point. We then decrease the $\lambda$ value and use the
estimate of $\bbeta$ from the last $\lambda$ as a warm start. The grid
of $\lambda$ can be chosen to be equally spaced on a log-scale, for
example, $\lambda_j=0.96^j\lambda_{\mathrm{max}}$ $(j=1,\ldots, m)$,
where $m$ is set so that
$\lambda_{\mathrm{min}}=0.2\lambda_{\mathrm{max}}$ or, alternatively,
we could terminate the loop until the model receives more than the
maximum number of nonzero coefficients allowed.

\section{Simulation studies} \label{secsim}
\subsection{Simulation strategies}
We simulate $n$ microbiome samples, $p$ nutrients and $q$ bacterial
taxa to mimic the real data set that we analyze in Section \ref
{secreal}. The nutrient intake vector is simulated using a
multivariate normal distribution with mean $\nulll$ and a covariance matrix
$\Sigma_{i,j}=\rho^{\vert i-j\vert}$. We simulate $p_r$ relevant
nutrients with each nutrient being associated with $q_r$ taxa. For each
nutrient, the association coefficients $\beta_{ij}$ for the $q_r$ taxa
are equally spaced over the interval $[0.6f, 0.9f]$ with alternative
signs, where $f$ controls the association strength.
We consider two growth models to relate the taxa abundances to the
covariates. In the exponential growth model, the proportion of the
$j$th taxon of the $i$th sample is determined as
%
\begin{equation}
\label{eqlinkmult} \phi_{ij} = \frac{ \exp (\beta_{j0} +\sum_{k=1}^p\beta_{jk}x_{ik} )}{ \sum_{j=1}^{q} \exp (\beta_{j0} +\sum_{k=1}^p\beta_{jk}x_{ik} )}.
\end{equation}
The intercepts $\bbeta_0$, which determine the base abundances of the
taxa, are taken from a uniform distribution over $(-2.3, 2.3)$ so that
the base taxa abundances can differ up to 100 folds. The exponential
growth model is a common model for bacteria growth in response to
environmental stimuli. We also consider a linear growth model, in which
the proportion of the $j$th taxon of the $i$th sample is determined as
\[
\phi_{ij} = \frac{\beta_{j0} + \sum_{k=1}^p\beta_{jk}x_{ik}}{ \sum_{j=1}^{q} (\beta_{j0} + \sum_{k=1}^p \beta_{jk}x_{ik})}.
\]
The intercepts $\bbeta_0$ are now drawn from a uniform distribution
over $(0.02, 2)$ so that the base taxa abundances can also differ up to
100 folds. To deal with possible negative $\sum_{k=0}^p\beta_{jk}x_{ik}$, we add a small constant to make it positive.

We then generate the count data using the DM model of parametrization
II (\ref{eqDM3}) with a common dispersion $\theta$. The number of
individuals (sequence reads) for the $i$th sample $m_i$ is generated
from a uniform distribution over $(m, 2m)$. Note that the data are not
generated exactly according to our model assumptions, which are based
on parametrization I (\ref{eqDM1}) and link (\ref{eqlink2}). This can
further demonstrate the robustness of our proposed model.

\begin{figure}[b]
\vspace*{-3pt}
\includegraphics{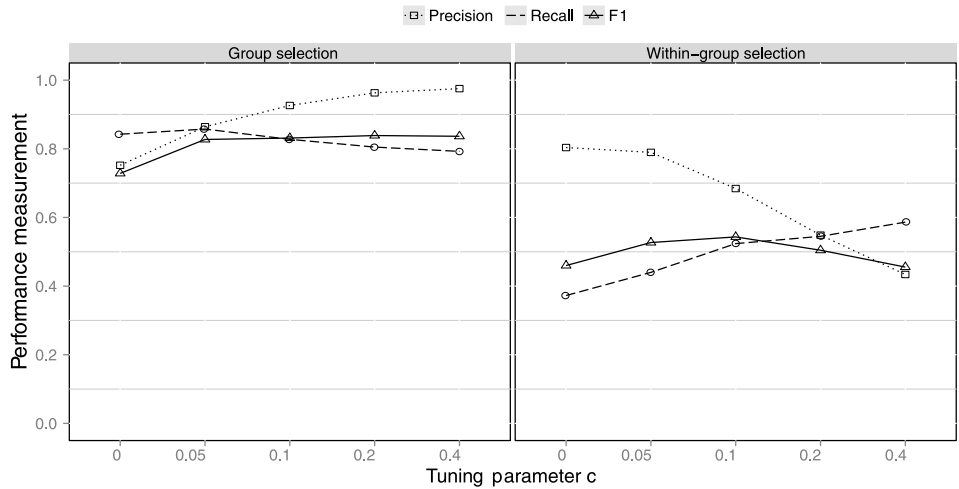}

\caption{{Effect of the tuning parameter $c$ on variable
selection.} The tuning parameter $c$ is varied from $0$ to $0.4$. Under
each value of $c$, the best $\lambda$ value, which maximizes the
likelihood of the test data set, is selected to generate the sparse
model. Group (left) and within-group (right) selection performances are
then evaluated using measures of recall, precision and $F_1$ based on
100 replications. Simulation setting: $n = 100$, $p =100$, $ p_r =
4$, $q = 40$, $q_r = 4$, $m = 500$, $\theta = 0.025$, $\rho = 0.4$.}
\label{figceffect}
\end{figure}

\subsection{Evaluation of the penalized likelihood approach for
selecting covariates affecting the microbiome composition}
To evaluate the variable selection performance of the proposed sparse
penalized likelihood approach with group $\ell_1$ penalty, we first
simulate the count data using the exponential growth model with $n =
100$, $p = 100$, $ p_r = 4$, $q = 40$, $q_r = 4$, $m = 1000$,
$\theta = 0.025$, and $\rho = 0.4$, totaling $4000$ variables. We compare
the results to the corresponding penalized estimation of the
DM model using only the $\ell_1$ penalty function and two other sparse
group $\ell_1$ estimations based on multinomial or Dirichlet
regression. In sparse multinomial regression, we use the multinomial
model for count data and\vadjust{\goodbreak} the link function is given by
(\ref{eqlinkmult}). We set $\beta_{10}=0$ to make the coefficients
identifiable. In sparse Dirichlet regression, instead of modeling the
counts directly, we model the proportions using the Dirichlet
distribution and the link function is the same as that of the DM
regression. Since the count data contain zeros, we add 0.5 to the cells
with 0 counts. We also include results from the LRT based univariate
testing procedure for group selection controlling the false discovery
rate (FDR) at 0.05.\looseness=-1

We measure the selection performance using
\[
\mathrm{recall} = \frac{\mathrm{TP}}{\mathrm{TP}+\mathrm{FN}},
\qquad \mathrm{precision} =
\frac
{\mathrm{TP}}{\mathrm{TP}+\mathrm{FP}},\qquad F_1 = 2 \cdot\frac{\mathrm{precision}
\cdot
\mathrm{recall}}{ \mathrm{precision} + \mathrm{recall}},
\]
where $\mathrm{TP}$, $\mathrm{FN}$ and $\mathrm{FP}$ are true positives, false negatives and false
positives, respectively, and $F_1$ is an overall measure, which weights
the precision and recall equally. The averages of these measures are
reported based on $100$ replications.

To select the best tuning parameter values, we simulate an independent
test data set of $n/2$ samples. We then run the penalized procedure
over the training data set and re-estimate the selected coefficients
using an unpenalized procedure (``nlm'' function in R). The
log-likelihood of the test data set is calculated based on the
re-estimated coefficients and the tuning parameter is selected to
maximize the log-likelihood over the test data set. We choose the
tuning parameter $c$ from the set $\{0, 0.05, 0.1, 0.2, 0.4\}$.
Figure~\ref{figceffect} shows that a small $c$ is sufficient to identify the
groups efficiently, while further increase of $c$ only improves the
group selection marginally. On the other hand, within-group selection
exhibits a unimode pattern indicating slight grouping could lead to
better identification of within-group elements. In the following
simulations, we tune both $c$ and $\lambda$ to achieve the maximum
likelihood values in the test data sets.


Table~\ref{tabsgl0} shows the simulation results. The sparse group
$\ell_1$ penalized DM regression has a much higher precision rate in
group selection than the corresponding $\ell_1$ penalized procedure,
while both achieve similar recall rates, demonstrating the gain from
including the group $\ell_1$ penalty in the regularization.
Interestingly, the sparse group penalized DM regression also performs
better in within-group selection, as shown by a higher recall rate and
$F_1$, indicating better group selection could also facilitate better
overall variable selection. Compared to models based on the sparse
Dirichlet regression and multinomial regression, the DM model performs
%
\begin{sidewaystable}
\textwidth=\textheight
\tablewidth=\textwidth
\caption{Comparison of sparse group $\ell_1$ and $\ell_1$ penalized
procedures for variable selection under Dirichlet-multinomial
(DM),\break
Dirichlet (D) and multinomial (M) regression models. The selection
performance, both group selection and\break within-group selection, is
evaluated using recall rate ({R}), precision rate ({P}) and $F_1$
({F}), all averaged over 100 runs\break (standard deviation in parenthesis).
The selection based on a univariate likelihood ratio test ({LRT}) at
FDR${}={}$0.05 is also indicated}\label{tabsgl0}
\begin{tabular*}{\tablewidth}{@{\extracolsep{\fill}}lcccccccccccc@{}}
\hline
& \multicolumn{6}{c}{\textbf{Sparse group} $\bolds{\ell_1}$ \textbf{penalization}} &
\multicolumn{6}{c@{}}{$\bolds{\ell_1}$ \textbf{penalization}} \\[-4pt]
& \multicolumn{6}{c}{\hrulefill}
& \multicolumn{6}{c@{}}{\hrulefill}\\
& \multicolumn{3}{c}{\textbf{Within-group}} & \multicolumn{3}{c}{\textbf{Group}} &
\multicolumn{3}{c}{\textbf{Within-group}} & \multicolumn{3}{c@{}}{\textbf{Group}}\\[-4pt]
& \multicolumn{3}{c}{\hrulefill} & \multicolumn{3}{c}{\hrulefill}
& \multicolumn{3}{c}{\hrulefill}
& \multicolumn{3}{c@{}}{\hrulefill}\\
\textbf{Model} & \textbf{R} & \textbf{P} & \textbf{F} & \textbf{R}
& \textbf{P} & \textbf{F} & \textbf{R} & \textbf{P} & \textbf{F} & \textbf{R}
& \textbf{P} & \textbf{F}\\
\hline\\[-8pt]
& \multicolumn{12}{c}{Exponential growth, $p = 100,q_r = 4, \theta
= 0.025$}\\[4pt]
DM & 0.59 & 0.70 & 0.59 & 0.86 & 0.92 & 0.87 & 0.42 &
0.76 & 0.48 & 0.88 & 0.68 & 0.70 \\
& (0.23) & (0.23) & (0.18) & (0.23) & (0.16) & (0.18) & (0.21) &
(0.23) & (0.18) & (0.22) & (0.29) & (0.22) \\
D & 0.48 & 0.73 & 0.52 & 0.83 & 0.89 & 0.82 & 0.36 &
0.82 & 0.45 & 0.82 & 0.77 & 0.72 \\
& (0.23) & (0.23) & (0.20) & (0.26) & (0.18) & (0.21) & (0.20) &
(0.21) & (0.19) & (0.26) & (0.27) & (0.23) \\
M & 0.46 & 0.72 & 0.50 & 0.82 & 0.85 & 0.79 & 0.36 &
0.76 & 0.44 & 0.84 & 0.70 & 0.69 \\
& (0.23) & (0.26) & (0.21) & (0.27) & (0.24) & (0.25) & (0.19) &
(0.24) & (0.18) & (0.26) & (0.28) & (0.24) \\
LRT & -- & -- & -- & 0.96 & 0.54 & 0.66 & -- & -- & -- &
0.96 & 0.54 & 0.66 \\
& -- & -- & -- & (0.11) & (0.21) & (0.16) & -- & -- & -- & (0.11) & (0.21) &
(0.16) \\
\hline
\end{tabular*}
\end{sidewaystable}
better in variable selection, especially for within-group selection,
suggesting the DM model is more appropriate than multinomial or
Dirichlet models when the counts exhibit overdispersion. The Dirichlet
model performs slightly better than the multinomial model. At 5\% FDR,
the LRT based univariate testing procedure selects far more variables
than these penalized procedures, yielding a higher recall rate but a
much worse precision rate.\looseness=1

\subsection{Effects of overdispersion and model misspecification}
\begin{figure}

\includegraphics{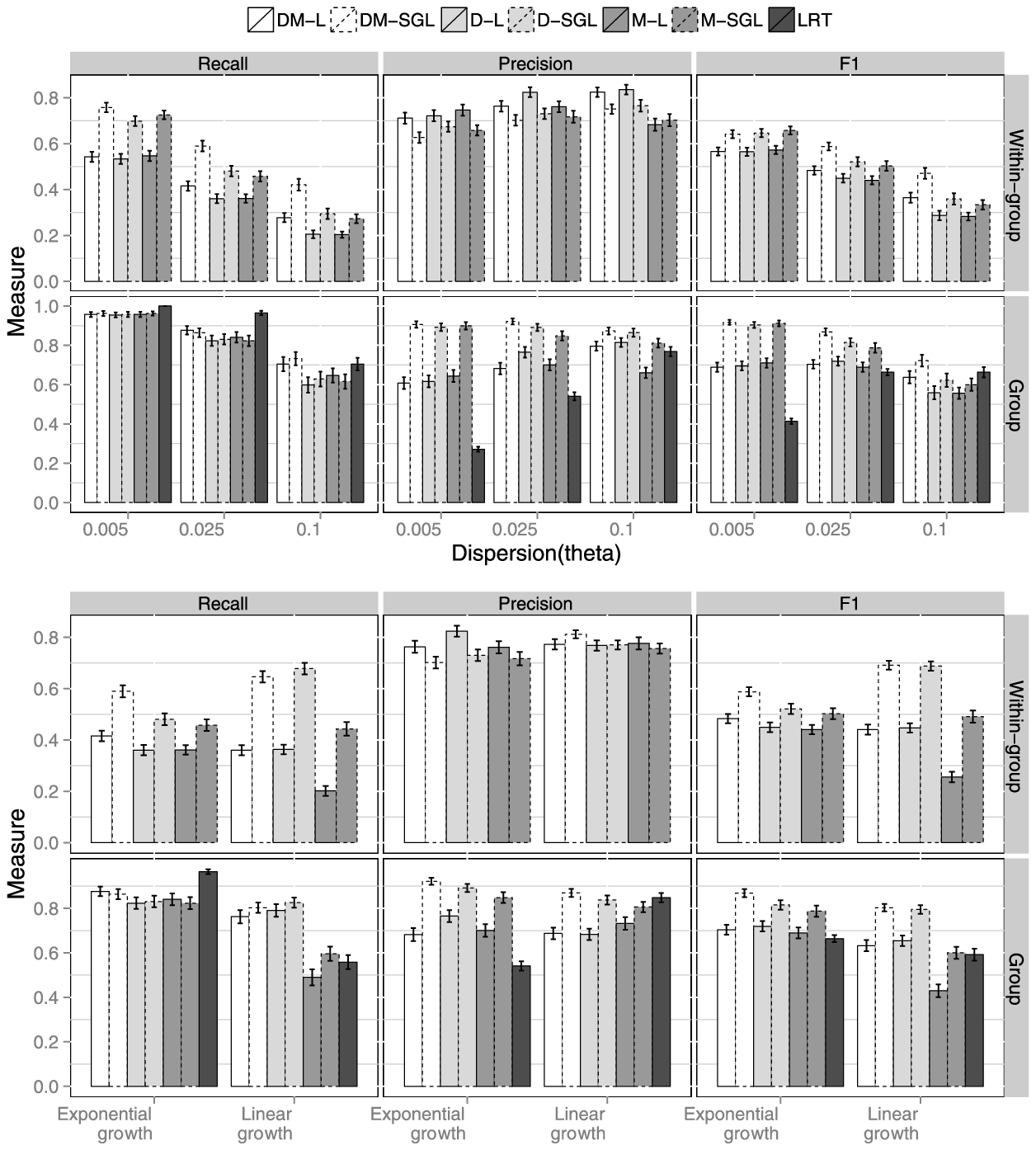}

\caption{Effects of overdispersion (top panel) and
model-misspecification (bottom panel) on the performance
of three different models and methods. DM-SGL: sparse group $\ell_1$
penalized Dirichlet-multinomial model;
DM-L: $\ell_1$ penalized Dirichlet-multinomial model; M-SGL: sparse
group $\ell_1$ penalized multinomial model; M-L: $\ell_1$ penalized
multinomial model; D-SGL: sparse group $\ell_1$ penalized Dirichlet model;
D-L: $\ell_1$ penalized Dirichlet model. For each bar,
mean${}\pm{}$standard error is presented based on 100 replications.}
\label{figsgl1}\vspace*{6pt}
\end{figure}

We further investigate the effect of overdispersion and simulate the
count data with different degrees of overdispersion and present the
results in Figure~\ref{figsgl1}. We observe that larger overdispersion
makes the selection more difficult for all three models, as shown by
smaller $F_1$ values. When the data have slight overdispersion ($\theta
 = 0.005$), the selection performances of the three models are
similar. On the other hand, when the data have much overdispersion
($\theta = 0.1$), DM performs much better than the other two models
in terms of both group selection and within-group selection. Therefore,
modeling overdispersion can lead to power gains in identifying relevant
variables if the data are overdispersed.

To assess the sensitivity to model misspecification, we simulate the
counts using the linear growth model instead and compare the results
with the exponential growth model (see Figure~\ref{figsgl1}).
Interestingly, both the Dirichlet and DM model are very robust to model
misspecification and their selection performances do not decrease
significantly. On the other hand, the multinomial model suffers a large
performance loss with the $F_1$ measure for group selection decreasing
from $0.79$ to $0.56$. We also study the effect of the total counts for
each sample (data not shown). Even increasing the total count by 10
folds, the DM model is still better than the proportion based Dirichlet
model. Therefore, even though we have much deeper sequencing of the
microbiome that results in larger counts for each sample, using the DM
model can still lead to improved performance over the model that
considers only the proportions.


\begin{figure}[b]

\includegraphics{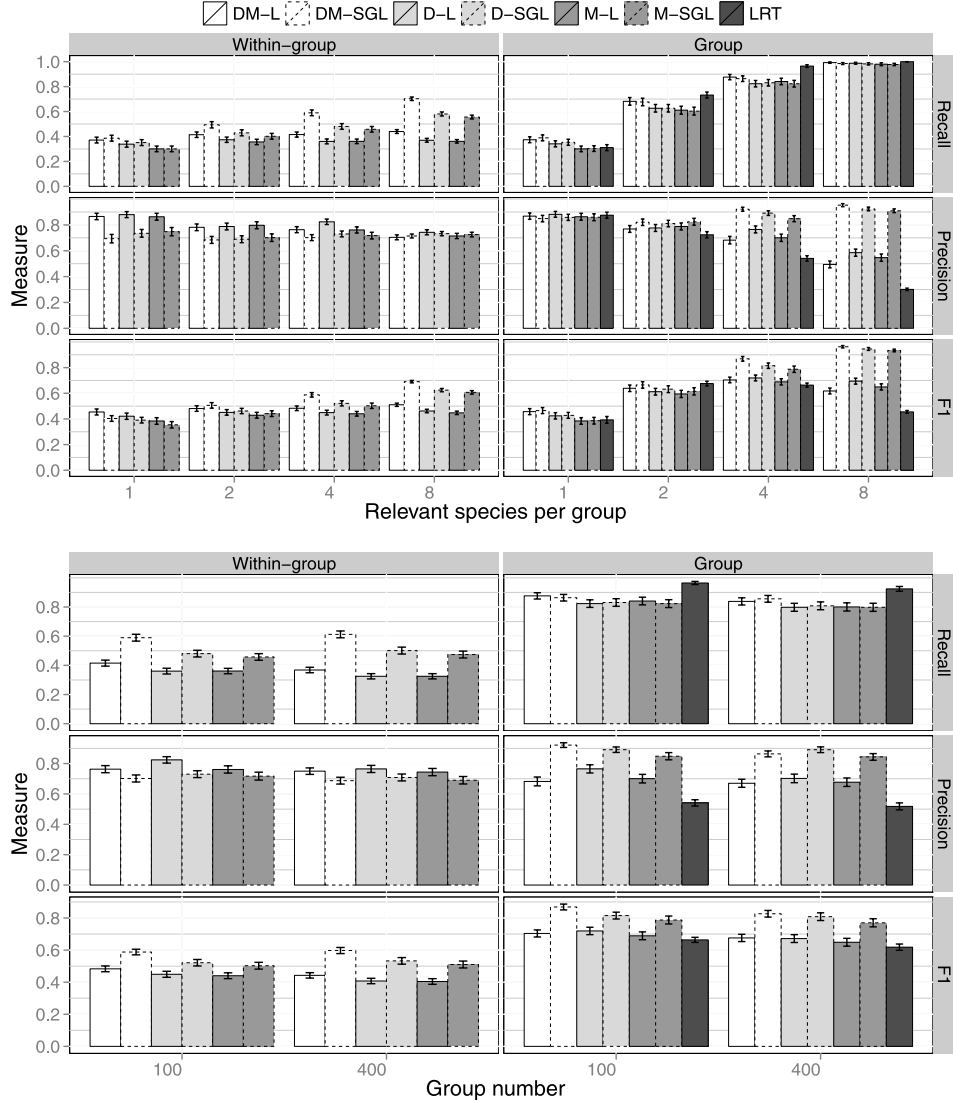}

\caption{Effects of the number of relevant taxa (top panel) and the
number of the covariates (bottom panel) on the performances of several
models and methods. DM-SGL: sparse group $\ell_1$ penalized
Dirichlet-multinomial model; DM-L: $\ell_1$ penalized
Dirichlet-multinomial model; M-SGL: sparse group $\ell_1$ penalized
multinomial model; M-L: $\ell_1$ penalized multinomial model; D-SGL:
sparse group $\ell_1$ penalized Dirichlet model; D-L: $\ell_1$
penalized Dirichlet model. For each bar, mean${}\pm{}$standard error is
presented based on 100 replications.} \label{figsgl3}
\end{figure}

\subsection{Effects of the number of the covariates and the relevant taxa}
We next study the effect of the number of relevant taxa in each group
on the performance of different models and present the results in
Figure~\ref{figsgl3}. When each relevant group contains only one
relevant taxon, the grouping is not very helpful, so the sparse group
regularized DM model and $\ell_1$ regularized DM model do not differ
much in selecting the relevant groups. When the relevant group contains
$8$ relevant taxa, variable grouping becomes much more important and
the sparse group regularized DM model performs much better than the
$\ell_1$ penalized DM. The group penalized multinomial and Dirichlet
regression models, on the other hand, select groups as well as the DM
regression model, since the grouping effect is much stronger.

Figure~\ref{figsgl3} also shows the results when we increase the
dimension of covariates to 400 ($16\mbox{,}000$ variables in total). Increase
of the dimension does not deteriorate the variable selection
performance, demonstrating the efficiency of our method in handling
high-dimensional data.

\section{Associating nutrient intake with the human gut microbiome
composition} \label{secreal}

Diet strongly affects the human health, partly by modulating gut
microbial community composition. \citet{Wu2011a} studied the habitual
diet effect on the human gut microbiome, where a cross-section of 98
healthy volunteers were enrolled in the study. Diet information was
collected using a food frequency questionnaire (FFQ) and was then
converted to nutrient intake values of 214 micronutrients. Nutrient
intake was further normalized using the residual method to adjust for
caloric intake and was standardized to have mean 0 and standard
deviation 1. Since some nutrient measurements were almost identical, we
used only one representative for these highly correlated nutrients
(correlation $\rho> 0.9$), resulting in 118 representative nutrients.
Stool samples were collected and DNA samples were analyzed by the
454$/$Roche pyrosequencing of 16S rDNA gene segments of the V1--V2 region.
The pyrosequences were denoised prior to taxonomic assignment, yielding
an average of $9265 \pm3864$ (SD) reads per sample. The denoised
sequences were then analyzed by the QIIME pipeline
[\citet{Caporaso2010a}] with the default parameter settings. The OTU table
contained 3068 OTUs (excluding the singletons) and these OTUs can be
further combined into 127 genera. We studied 30 relatively common
genera that appeared in at least 25 subjects. Finally, we had the count
matrix $\Y_{98\times30}$ and covariate matrix $X_{98\times118}$. Our
goal is to identify the micronutrients that are associated with the gut
microbiomes and the specific genera that the selected nutrients affect.

\begin{figure}

\includegraphics{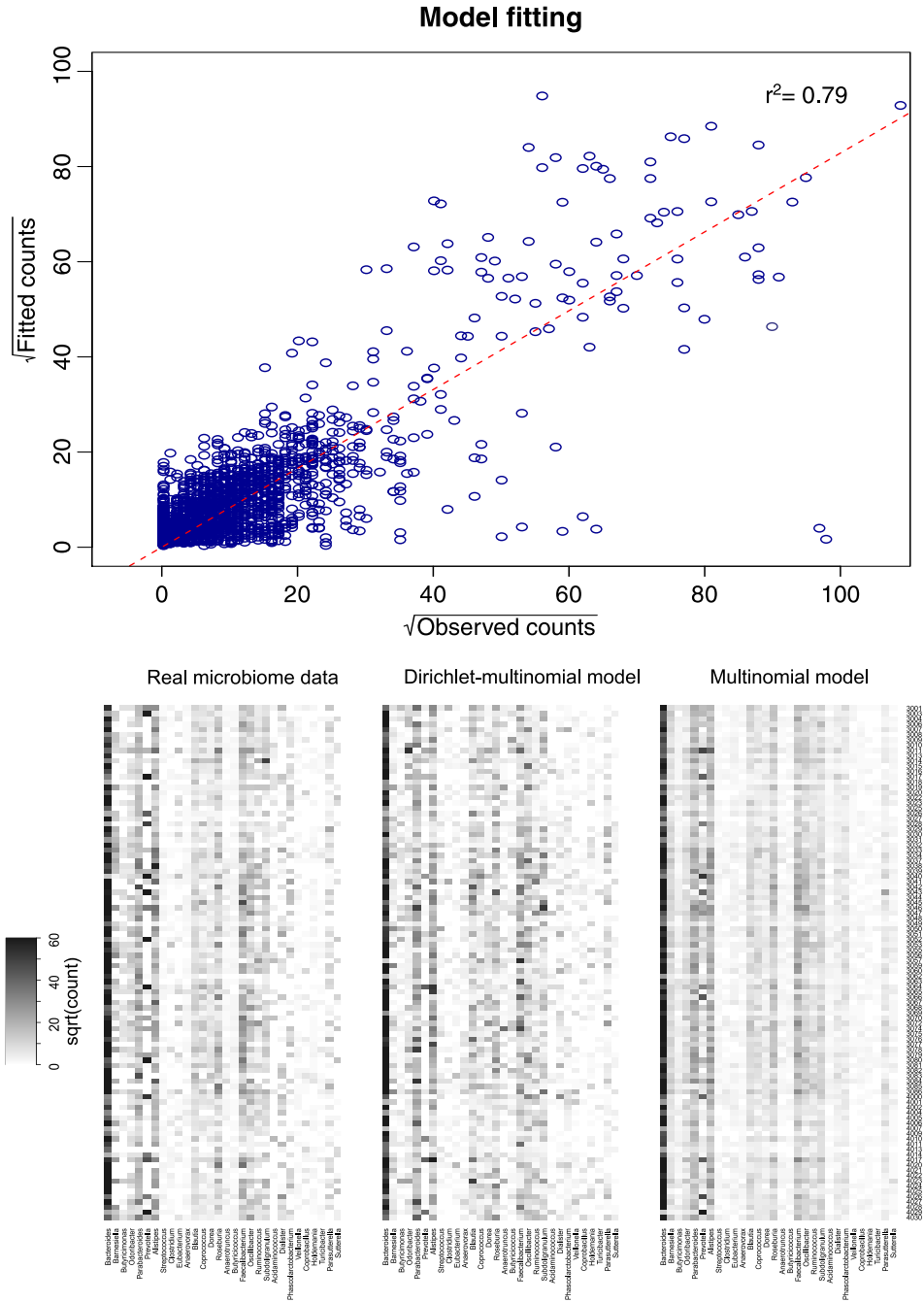}

\caption{Model fit using the variables selected by the sparse group
$l_1$ penalized DM model. Top plot: square root of the fitted counts
versus square root of the observed counts based on the DM model with
the selected nutrients; bottom plots: observed counts and simulated
counts produced by the fitted sparse DM model and multinomial model.}
\label{figfit}
\end{figure}

\begin{sidewaystable}
\tabcolsep=0pt
\textwidth=\textheight
\tablewidth=\textwidth
\caption{Estimated regression coefficients from the sparse group
$\ell_1$ penalized DM regression for the diet-gut microbiome data.\break The
exponentiation of a given coefficient can be interpreted as the factor
of change in proportion of a taxon when\break a given nutrient changes by one
unit while other nutrients remain constant. Columns 1--11 represent the
selected nutrients: Polyunsaturated fat, Methionine, Sucrose, Animal
Protein, Vitamin E-Food Fortification, Maltose, Added Germ from
wheats, Choline-Phosphatidylcholine, Taurine, Naringenin-flavanone and
Eriodictyol-flavonone. Rows 1--13 represent the selected\break bacteria taxa:
Bacteroides, Barnesiella, Odoribacter, Parabacteroides, Prevotella,
Alistipes, Coprococcus, Faecalibacterium,\break Oscillibacter, Ruminococcus,
Subdoligranulum, Phascolarctobacterium and Parasutterella. The marginal
$p$-value\break based on the LRT and the bootstrap selection probability of
each of the selected nutrients are also shown}\label{estcoeff}
\begin{tabular*}{\tablewidth}{@{\extracolsep{4in minus 4in}}ld{2.2}d{2.2}cd{2.2}d{2.2}d{2.2}
d{2.2}d{2.2}d{2.2}d{2.2}@{}}
\hline
\multicolumn{11}{c}{\textbf{Row: taxon; column: nutrient}}\\
\hline
-- &-0.03&-0.08 &0.09 &-0.08 &-0.10 &-0.02 &0.02 &0.10 & \multicolumn{1}{c}{--} &-0.03 \\
$-$0.32 & \multicolumn{1}{c}{--} &-0.33 & \multicolumn{1}{c}{--} & \multicolumn{1}{c}{--} & \multicolumn{1}{c}{--} &0.22 & \multicolumn{1}{c}{--} & \multicolumn{1}{c}{--} & \multicolumn{1}{c}{--} & \multicolumn{1}{c@{}}{--} \\
$-$0.38 & \multicolumn{1}{c}{--} & \multicolumn{1}{c}{--} & \multicolumn{1}{c}{--} & \multicolumn{1}{c}{--} & \multicolumn{1}{c}{--} & \multicolumn{1}{c}{--} & \multicolumn{1}{c}{--} & \multicolumn{1}{c}{--} &-0.29 & \multicolumn{1}{c@{}}{--} \\
-- &-0.01 &-0.08 &0.13 &-0.07 & \multicolumn{1}{c}{--} & \multicolumn{1}{c}{--} & \multicolumn{1}{c}{--} &0.02 &-0.23 & \multicolumn{1}{c@{}}{--} \\
-- & \multicolumn{1}{c}{--} &0.23 & \multicolumn{1}{c}{--} & \multicolumn{1}{c}{--} &0.36 &0.63 &-0.72 & \multicolumn{1}{c}{--} & \multicolumn{1}{c}{--} & \multicolumn{1}{c@{}}{--} \\
$-$0.19 &-0.04 & \multicolumn{1}{c}{--} &0.16 & \multicolumn{1}{c}{--} & \multicolumn{1}{c}{--} & \multicolumn{1}{c}{--} & \multicolumn{1}{c}{--} & \multicolumn{1}{c}{--} & \multicolumn{1}{c}{--} &0.05 \\
-- & \multicolumn{1}{c}{--} & \multicolumn{1}{c}{--} & \multicolumn{1}{c}{--} & \multicolumn{1}{c}{--} & \multicolumn{1}{c}{--} & \multicolumn{1}{c}{--} & \multicolumn{1}{c}{--} & \multicolumn{1}{c}{--} & \multicolumn{1}{c}{--} &0.16 \\
-- & \multicolumn{1}{c}{--} & \multicolumn{1}{c}{--} & \multicolumn{1}{c}{--} & \multicolumn{1}{c}{--} &-0.08 & \multicolumn{1}{c}{--} & \multicolumn{1}{c}{--} & \multicolumn{1}{c}{--} &0.07 & \multicolumn{1}{c@{}}{--} \\
-- &-0.02 & \multicolumn{1}{c}{--} & \multicolumn{1}{c}{--} & \multicolumn{1}{c}{--} & \multicolumn{1}{c}{--} & \multicolumn{1}{c}{--} & \multicolumn{1}{c}{--} &-0.10 & \multicolumn{1}{c}{--} & \multicolumn{1}{c@{}}{--} \\
-- & \multicolumn{1}{c}{--} &0.19 & \multicolumn{1}{c}{--} & \multicolumn{1}{c}{--} & \multicolumn{1}{c}{--} & \multicolumn{1}{c}{--} & \multicolumn{1}{c}{--} & \multicolumn{1}{c}{--} & \multicolumn{1}{c}{--} & \multicolumn{1}{c@{}}{--} \\
-- &0.02 & \multicolumn{1}{c}{--} & \multicolumn{1}{c}{--} & \multicolumn{1}{c}{--} & \multicolumn{1}{c}{--} & \multicolumn{1}{c}{--} & \multicolumn{1}{c}{--} &-0.12 &-0.12 &0.14 \\
-- & \multicolumn{1}{c}{--} & \multicolumn{1}{c}{--} & \multicolumn{1}{c}{--} &-0.35 & \multicolumn{1}{c}{--} & \multicolumn{1}{c}{--} & \multicolumn{1}{c}{--} & \multicolumn{1}{c}{--} & \multicolumn{1}{c}{--} & \multicolumn{1}{c@{}}{--} \\
$-$0.26 & \multicolumn{1}{c}{--} &-0.29 & \multicolumn{1}{c}{--} & \multicolumn{1}{c}{--} & \multicolumn{1}{c}{--} & \multicolumn{1}{c}{--} & \multicolumn{1}{c}{--} & \multicolumn{1}{c}{--} & \multicolumn{1}{c}{--} & \multicolumn{1}{c@{}}{--} \\
\hline
\multicolumn{11}{c}{\textbf{Marginal} $\bolds{p}$\textbf{-value}}\\
\hline
$4.5\times10^{-3}$ & \multicolumn{1}{c}{$2.2\times10^{-4}$}
& \multicolumn{1}{c}{$8.4\times10^{-4}$} &
\multicolumn{1}{c}{$3.6\times10^{-4}$}
& \multicolumn{1}{c}{$1.1\times10^{-1}$} & \multicolumn{1}{c}{$6.0\times10^{-3}$} &
\multicolumn{1}{c}{$9.5\times10^{-6}$}& \multicolumn{1}{c}{$2.7\times10^{-3}$}
& \multicolumn{1}{c}{$5.9\times10^{-3}$} & \multicolumn{1}{c}{$5.8\times10^{-2}$}
& \multicolumn{1}{c@{}}{$5.2\times10^{-3}$}\\
\hline
\multicolumn{11}{c}{\textbf{Bootstrap selection probability}}\\
\hline
0.50 & 0.93 & 0.72 &0.94&0.35&0.67 & 0.43& 0.58& 0.92& 0.61&0.60\\
\hline
\end{tabular*}
\end{sidewaystable}

We applied the sparse group $\ell_1$ penalized DM regression to this
data set. We used the BIC to select the tuning parameters. The final DM
model selected 11~nutrients and 13 associated genera. We refit the DM
regression model using the selected variables and obtained the maximum
likelihood estimates of the coefficients. We compared the fitted counts
(total count${}\times{}$fitted proportion) against the observed counts in
Figure~\ref{figfit} (top panel). The model fits the data quite well
with $r^2=0.79$. Table~\ref{estcoeff} shows the MLEs of the regression
coefficients for the selected nutrients and genera.
Except for Methionine (second column), the coefficients are not too
small.\vadjust{\goodbreak} Since the nutrient measurements are standardized, the
exponentiation of a given coefficient can be interpreted as the factor
of change in proportion of a taxon when a given nutrient changes by one
unit while other nutrients remain constant. The marginal $p$-value
based on the LRT for each of the selected nutrients is also shown in
this table. Except for Vitamin E and Eriodictyol, these selected
nutrients all showed a significant marginal association with the gut microbiome.

To further assess the relevance of the nutrients selected, we used the
bootstrap to analyze the stability of the selected nutrients
[\citet{bach08}]. Specifically, we took 100 bootstrap samples and for each
sample we ran our algorithm to select the nutrients. Since some
nutrients are highly correlated, we expect that highly correlated
nutrients (if the correlation is greater than 0.75) can be selected in
different bootstrap samples; we define the bootstrap selection
probability of a given nutrient as the number of times that this
nutrient or its correlated nutrients were selected. Table \ref
{estcoeff} shows the bootstrap probabilities of the nutrients that
were selected by the sparse DM regression, indicating quite stable
selection of most of the selected microbiome-associated nutrients.
Vitamin E had the least stable selection over the 100 bootstrap samples.

\begin{figure}

\includegraphics{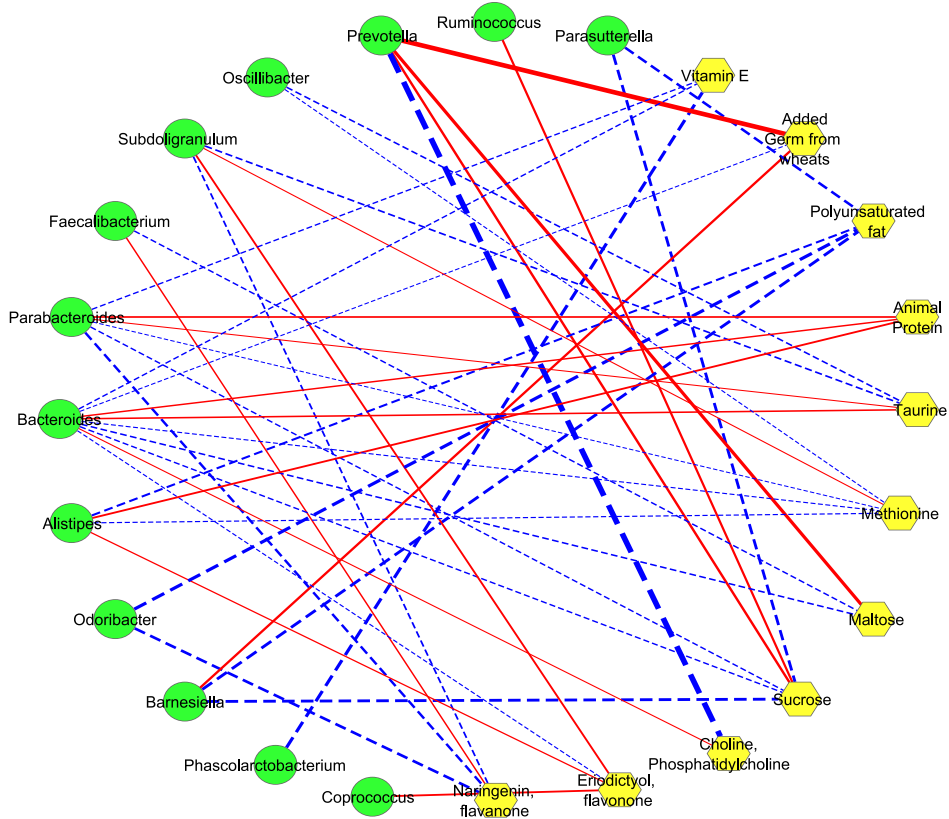}

\caption{Association of nutrients with human gut microbial taxa
identified by the sparse group $\ell_1$ regularized DM model. We use a
bipartite graph to visualize the selected nutrients and their
associated genera based on sparse group $\ell_1$ penalized DM
regression. Circle: genus; hexagon: nutrient; solid line: positive
correlation; dashed line: negative correlation. The thickness of the
line represents the association strength.} \label{figcombo}
\end{figure}

The identified nutrient-taxon associations are visualized in a
bipartite graph shown in Figure~\ref{figcombo}, where the genera and
nutrients are depicted with circles and hexagons, respectively. These
results further confirmed the findings of \citet{Wu2011a}, where they
found the human gut microbiome can be clustered into two enterotypes
characterized by Prevotella and Bacteroides, respectively, and the
Prevotella enterotype is associated with a high carbohydrate diet while
the Bacteroides enterotype is associated with a high
protein/fat/choline diet. Figure~\ref{figcombo} shows that two
carbohydrates, Maltose and Sucrose, are positively associated with
Prevotella and negatively associated with Bacteroides, while animal
proteins are positively associated with Bacteroides, Parabacteroides
and Alistipes, the three genera mostly enriched in the Bacteroides
enterotype. Choline is positively associated with Bacteroides and
negatively associated with Prevotella. Polyunsaturated fat is strongly
associated with Alistipes, Odoribacter, Barnesiella and Parasutterella,
indicating the large effect of fat on the human microbiome.

The DM model also identified several other associations that are worth
further investigation. For example, we found that Naringenin
(flavanone) was positively associated with Faecalibacterium, an
anti-inflammatory commensal bacterium identified by gut microbiota
analysis of Crohn's disease patients [\citet{Sokol2008}]. If the
association is validated, diet with high Naringenin (e.g., Orange,
Grapefruit) can be beneficial for patients with Crohn's disease.


As a comparison, we also ran the sparse group $\ell_1$ penalized
multinomial or Dirichlet regression models and the identified
nutrient-genus associations showed significant overlap with those from
the DM regression model. However, the interpretability of the DM
regression model was the best. To further demonstrate the advantage of
the DM model, we simulated taxa counts for each individual based on the
fitted models and the observed total taxa counts. The bottom plot of
Figure~\ref{figfit} shows that the simulated counts produced by the
fitted sparse DM model resemble the observed counts better than those
from the sparse multinomial model, where the simulated counts are
apparently over-smoothed. This indicates the
importance of considering the overdispersion in modeling the gut
microbiome data.
We also performed the LRT based univariate testing procedure. At
FDR${}={}$0.05, the LRT identified 13 nutrients, 8 of which are also
identified or highly correlated with the nutrients identified by the
sparse group $\ell_1$ penalized DM model.

\section{Discussion} \label{secdis}
We have proposed a sparse group $\ell_1$ penalized estimation for the
DM regression in order to select covariates associated with the
microbiome composition. The sparse group $\ell_1$ penalty encourages
both group-level and within-group sparsity, with which we can select
the relevant taxa associated with the selected covariates. We have
performed extensive simulations to evaluate our proposed penalized
estimation procedure for both group and within-group selections. We
demonstrated the procedure with a real data set on associating nutrient
intakes with gut microbiome composition and confirmed the major
findings in \citet{Wu2011a}.

In our penalized likelihood estimation of the DM model, we use a
combination of group $\ell_1$ and individual $\ell_1$ penalties, which
result in a convex and separable (in groups of parameters) penalty
function. This property facilitates the application of the general
coordinate gradient descent method of \citet{Tseng2007} to
implement an efficient optimization algorithm. In each iteration, we
have a closed form solution for a block update. For a given set of the
sparsity tuning parameters, our algorithm is fully automatic and does
not require the specification of an algorithmic tuning parameter to
ensure convergence. For example, it took about 3 minutes on a standard
laptop (Core i5, 2G memory) to finish the analysis of the real data set
using an R implementation of the algorithm (available at
\url{http://statgene.med.upenn.edu/}). Besides the sparse $l_1$ group
penalty, other group penalty functions such as the sup-norm penalty in
\citet{zhang-sup} and the composite absolute penalties in
\citet{zhao09} can also be used in the setup of the Dirichlet
multinomial regression. However, efficient implementation of the
optimization problems with these penalty functions is challenging.

In microbiome data analysis literature, one commonly used approach is
to normalize the counts into proportions and perform statistical
analysis using the proportions. However, by converting into the
proportions, the variation associated with the multinomial sampling
process is lost. In 16S rRNA sequencing, the sequencing depths (total
counts) for samples can vary up to 10-fold. Obviously, the accuracy of
the proportion estimates under sequencing depth of $500$ reads is very
different from that of $10\mbox{,}000$ reads. As shown in our simulations,
modeling counts directly can result in gain of power in selecting
relevant variables even when the number of sequence reads is very
large. Another problem associated with proportions is the existence of
numerous zeros in the taxa count data. Many proportion based approaches
require taking logarithms of the proportions, which is problematic for
the zero proportions. To circumvent this problem, either a pseudo count
(e.g., 0.5) is added to these zero counts before converting into
proportions or an arbitrary small proportion is substituted for these
zero proportions. The effects of creating pseudo counts have not been
evaluated thoroughly when the data contain excessive zeros.

Besides overdispersion, the taxa count data can also exhibit
zero-inflation [\citet{BARRY2002}], where the count data contain
more zeros than expected from the DM model. How to model the microbiome
count data that allows overdispersion, zero-inflation and possibly the
phylogenetic correlations among the taxa is an important future
research topic. The multilevel zero-inflated DM regression model for
overdispersed count data with extra zeros
[\citet{Moghimbeigi2008,Lee2006}] can potentially provide a
solution to this problem. Another problem associated with the DM model
is its inflexibility in modeling the covariance structure among the
taxa counts. The multinomial model for counts compounded by a logistic
normal model [\citet{aitchison82}] for proportions provides a
possible solution. This needs to be investigated further.

\begin{appendix}\label{app}
\section*{Appendix}
%
\begin{theorem} \label{theorem1}
Letting $\b,{\mathbf{x}}\in\reals^n$, $\lambda_1,\lambda_2,c$ are
nonnegative
constants and ${\mathbf{x}}^0$ is the minimizer of the following
function:
%
\begin{equation}
\label{min1} f({\mathbf{x}}) = \tfrac{1}{2}{\mathbf{x}}^T{
\mathbf{x}}+ \b^T{\mathbf{x}}+ c + \lambda_1 \llVert {
\mathbf{x}} \rrVert_2 + \lambda_2 \llVert {\mathbf{x}}
\rrVert_1,
\end{equation}
then ${\mathbf{x}}^0_S = \nulll$ and
\[
{\mathbf{x}}^0_{\bar{S}} = \arg\min_{{\mathbf{x}}_{\bar{S}}} \bigl\{
\tfrac{1}{2}{\mathbf{x}}^T_{\bar
{S}}{\mathbf{x}}_{\bar{S}}
+ \bigl(\b_{\bar{S}} - \lambda_2 \operatorname{sgn}(
\b_{\bar
{S}}) \bigr)^T{\mathbf{x}}_{\bar{S}} + c +
\lambda_1 \llVert {\mathbf{x}}_{\bar{S}} \rrVert_2
\bigr\},
\]
where $S=\{i \in\{1, \ldots,n \} \vert \vert b_i \vert< \lambda_2
\}$ and $\bar{S} = \{1, \ldots,n \} \setminus S$ and
$\operatorname{sgn}(\cdot)$ is the sign function.
\end{theorem}
\begin{pf}
We prove ${\mathbf{x}}^0_S = \nulll$ by contradiction. If $x^0_i
\neq0$ $(i \in S)$,
then we can construct a new ${\mathbf{x}}^1$ with $x^1_i=0$ and other
components
being the same as ${\mathbf{x}}^0$. Clearly, $\frac{1}{2}{{\mathbf
{x}}^1}^T{\mathbf{x}}^1 + \b^T{\mathbf{x}}^1
+ c + \lambda_2 \llVert {\mathbf{x}}^1 \rrVert_1 < \frac
{1}{2}{{\mathbf
{x}}^0}^T{\mathbf{x}}^0 + \b^T{\mathbf{x}}^0 + c +
\lambda_2 \llVert {\mathbf{x}}^0 \rrVert_1$
and $\lambda_1 \llVert {\mathbf{x}}^1 \rrVert_2 < \lambda_1 \llVert {\mathbf{x}}^0 \rrVert_2$. The former is
due to the fact that
$\frac{1}{2}(x_i^0)^2 + b_ix_i^0 + \lambda_2 \vert x_i^0 \vert> 0$
for $\vert b_i \vert< \lambda_2$. Hence, ${\mathbf{x}}^0$ is not the
minimizer
of $f({\mathbf{x}})$, which is contradictory. Therefore, ${\mathbf
{x}}^0_S = \nulll$.

To prove the second part, we note that $x_i^0$ must be either $0$ or
have an opposite sign of $b_i$ for $i \in\{1,\ldots, n\}$. So the
minimization of $\f({\mathbf{x}})$ is equivalent to minimizing
\[
f^*({\mathbf{x}}) = \tfrac{1}{2}{\mathbf{x}}^T{\mathbf{x}}+
\bigl(\b- \lambda_2\operatorname{sgn}(\b)\bigr)^T{
\mathbf{x}}+ c + \lambda_1 \llVert {\mathbf{x}} \rrVert_2,
\]
subject to
\[
\operatorname{sgn}(x_i) = -\operatorname{sgn}(b_i)
\quad\mbox{or}\quad  x_i=0.
\]
Since ${\mathbf{x}}^0_S = \nulll$, we can restrict the minimization
over only ${\mathbf{x}}_{\bar{S}}$,
%
\begin{equation}
\label{min2} f^*({\mathbf{x}}_{\bar{S}}) = \tfrac{1}{2}{
\mathbf{x}}_{\bar
{S}}^T{\mathbf{x}}_{\bar{S}} + \bigl(
\b_{\bar
{S}}- \lambda_2\operatorname{sgn}(\b_{\bar{S}})
\bigr)^T{\mathbf{x}}_{\bar
{S}} + c + \lambda_1
\llVert {\mathbf{x}}_{\bar{S}} \rrVert_2,
\end{equation}
subject to
\[
\operatorname{sgn}(x_i) = -\operatorname{sgn}(b_i)
\quad\mbox{or}\quad x_i=0\qquad (i \in \bar{S}).
\]
Since ${\mathbf{x}}^0_{\bar{S}}$ is the minimizer of $f^*({\mathbf
{x}}_{\bar{S}})$ without
the constraint, the sign of ${\mathbf{x}}^0_{\bar{S}}$ should be the
opposite of
the sign of $(\b_{\bar{S}}- \lambda_2\operatorname{sgn}(\b_{\bar{S}}))$.
Because $\vert b_i \vert\ge\lambda_2$ for $i \in\bar{S}$, the sign
of $(\b_{\bar{S}}- \lambda_2\operatorname{sgn}(\b_{\bar{S}}))$ is the
same as
$\b_{\bar{S}}$. So the sign of ${\mathbf{x}}^0_{\bar{S}}$ is the
opposite of that
of $\b_{\bar{S}}$. Therefore, ${\mathbf{x}}^0_{\bar{S}}$ satisfies
the constraint.\vspace*{-2pt}
\end{pf}

Using simple variable substitution, we have the following corollary.
%
\begin{corollary} \label{cor1}
Letting $\b,\bbeta,\d\in\reals^n$, $\lambda_1,\lambda_2,c$ are
nonnegative constants and $\d^0$ is the minimizer of the following function,
%
\begin{equation}
f(\d) = \tfrac{1}{2}\d^T\d+ \b^T\d+ c +
\lambda_1 \llVert \bbeta+\d\rrVert_2 +
\lambda_2 \llVert \bbeta+\d\rrVert_1,
\end{equation}
then $\d^0_S = -\bbeta_S$ and
\[
\d^0_{\bar{S}} = \arg\min_{\d_{\bar{S}}} \biggl\{
\frac{1}{2}\d^T_{\bar
{S}}\d_{\bar{S}} + \bigl(
\b_{\bar{S}} - \lambda_2 \operatorname{sgn}(\b_{\bar
{S}}-
\bbeta_{\bar{S}}) \bigr)^T\d_{\bar{S}} + c +
\lambda_1 \llVert \d_{\bar{S}}+\bbeta_{\bar{S}}
\rrVert_2 \biggr\},
\]
where $S=\{i \in\{1, \ldots,n \} \vert \vert b_i - \beta_i\vert<
\lambda_2 \}$, $\bar{S} = \{1, \ldots,n \} \setminus S$ and
$\operatorname{sgn}(\cdot)$ is the sign function.\vspace*{-2pt}
\end{corollary}
\end{appendix}

\section*{Acknowledgments}

We thank Doctors Rick Bushman, James Lewis and Gary Wu for providing
the data and for many insightful discussions. We also thank Professor
Karen Kafadar, an Associate Editor and two reviewers for many helpful
comments.



\printaddresses

\end{document}